\documentclass[twocolumn,aip,reprint,numerical]{revtex4-2}
\usepackage{array}
\usepackage{float}
\usepackage{textcomp}
\usepackage{tipa}
\usepackage{amsmath}
\usepackage{amssymb}
\usepackage{graphicx}
\usepackage{hyperref}
\hypersetup{hypertex=true,
            colorlinks=true,
            linkcolor=blue,
            anchorcolor=blue,
            citecolor=blue}
\usepackage{array}
\usepackage{amsfonts}
\usepackage{amssymb,amsmath,multirow,rotate}
\usepackage{mathrsfs}
\usepackage{booktabs}
\usepackage{multirow}
\usepackage{threeparttable}
\usepackage{chngpage}
\usepackage{latexsym}
\usepackage{dcolumn}
\usepackage{graphics,graphicx,bm,fleqn,epic,eepic,float}
\usepackage{times}
\usepackage{verbatim}
\usepackage{subfig}
\usepackage{color}
\usepackage{xr}
\usepackage{listings} 
\usepackage{placeins} 
\usepackage{tikz}
\usepackage{url}
\usepackage{soul}
\usepackage{caption}
\usepackage{natbib}

\usepackage{amssymb}
\usepackage{pifont}
\newcommand{\cmark}{\ding{51}}%
\newcommand{\xmark}{\ding{55}}%
\definecolor{red}{rgb}{1,0,0}

\definecolor{blue}{rgb}{0,0,1}

\makeatletter

\newcommand{\lyxmathsym}[1]{\ifmmode\begingroup\def\b@ld{bold}
        \text{\ifx\math@version\b@ld\bfseries\fi#1}\endgroup\else#1\fi}

\usepackage{dcolumn}
\usepackage{epsfig}
\usepackage{braket}

\usepackage{lineno}
\makeatother

\begin{document}

\title{Random forests for detecting weak signals and extracting physical information: a case study of magnetic navigation}

\date{\today}

\author{Mohammadamin Moradi}
\affiliation{School of Electrical, Computer and Energy Engineering, Arizona State University, Tempe, AZ 85287, USA}

\author{Zheng-Meng Zhai}
\affiliation{School of Electrical, Computer and Energy Engineering, Arizona State University, Tempe, AZ 85287, USA}

\author{Aaron Nielsen}
\affiliation{ANT Center, Air Force Institute of Technology, 2950 Hobson Way, Wright-Patterson AFB, OH 45433, USA}

\author{Ying-Cheng Lai} \email{Ying-Cheng.Lai@asu.edu}
\affiliation{School of Electrical, Computer and Energy Engineering, Arizona State University, Tempe, AZ 85287, USA}
\affiliation{Department of Physics, Arizona State University, Tempe, Arizona 85287, USA}

\begin{abstract}

It was recently demonstrated that two machine-learning architectures, reservoir computing and time-delayed feed-forward neural networks, can be exploited for detecting the Earth's anomaly magnetic field immersed in overwhelming complex signals for magnetic navigation in a GPS-denied environment. The accuracy of the detected anomaly field corresponds to a positioning accuracy in the range of 10 to 40 meters. To increase the accuracy and reduce the uncertainty of weak signal detection as well as to directly obtain the position information, we exploit the machine-learning model of random forests that combines the output of multiple decision trees to give optimal values of the physical quantities of interest. In particular, from time-series data gathered from the cockpit of a flying airplane during various maneuvering stages, where strong background complex signals are caused by other elements of the Earth's magnetic field and the fields produced by the electronic systems in the cockpit, we demonstrate that the random-forest algorithm performs remarkably well in detecting the weak anomaly field and in filtering the position of the aircraft. With the aid of the conventional inertial navigation system, the positioning error can be reduced to less than 10 meters. We also find that, contrary to the conventional wisdom, the classic Tolles-Lawson model for calibrating and removing the magnetic field generated by the body of the aircraft is not necessary and may even be detrimental for the success of the random-forest method.

\end{abstract}

\maketitle

\section{Introduction} \label{sec:intro}

Random forest~\cite{breiman2001random,rigatti2017random,cutler2012random} is
a supervised machine-learning method for solving classification and regression
problems based on noisy feature data. The ``forest'' consists of a number
decision trees, each trained using a different subset of the characteristics
and data points. A decision tree~\cite{su2006fast,freund1999alternating}
recursively splits the feature space into two halves, which is done for a
specific feature input using a threshold and the depth of the tree is the
number of available input feature signals. At the end of the tree construction,
the whole feature space has been divided into a large number of small subspaces,
each associated with a particular value of the physical quantity of interest
(the target variable), leading to a ``leaf.'' Given a number of feature signals
in the form of, e.g., time series, and the value of the target variable,
supervised training can be done through some standard optimization method to
determine a proper set of the threshold values required for splitting the feature 
signals. After training is done, when a new set of feature signals is presented 
to the tree, a branch of the tree can be quickly identified which closely 
matches the corresponding feature values, and the leaf end of this branch gives 
the predicted value of the physical quantity for the particular set of feature
values. In principle, a decision tree can be used for predicting the values
of the target variable corresponding to different combinations of the values
of the feature signals, but a single tree is susceptible to overfitting, 
especially when the feature space becomes large. Random forest solves this
overfitting problem by combining  a number of random decision trees, each 
responsible for a subset of the features and a portion of the training data. 

More specifically, the working of the random-forest algorithm can be described, 
as follows. For each tree in the forest, a subset of the features and a portion 
of the training data are randomly chosen, where each tree is given a slightly
different perspective on the data, known as as bootstrap
sampling~\cite{hesterberg2011bootstrap}. The algorithm then creates a decision
tree utilizing the characteristics and data that have been chosen, using a
splitting criterion such as the Gini index or information gain. The new tree is
then added to the forest, and the process is repeated until a pre-specified
requirement is met, at which point no more trees are produced. During the
testing or prediction phase, the model combines the predictions from all the
trees in the forest to provide a forecast. The predicted value of the
target variable is the average or median of the predictions of all trees in
the ensemble. Random forest is resilient to noise and outliers, and it
alleviates overfitting since the ensemble of trees balances out the noise and
variability in the data. Moreover, random forests is capable of handling
missing data and can offer metrics of feature relevance. Random forest 
regression has been applied in various fields, such as
healthcare~\cite{simsekler2020evaluation,iwendi2020covid} and
transportation~\cite{manzoor2018vehicle, urbancic2018transportation}, for
tasks such as predicting traffic flow~\cite{zhang2018hybrid} and forecasting
commodity prices~\cite{ma2019stock}. Various extensions and modifications of
the random-forest algorithm, such as extremely randomized
trees~\cite{geurts2006extremely} and quantile regression
forests~\cite{meinshausen2006quantile}, have been proposed to further improve
the accuracy and robustness of the method.

In this paper, we exploit random forests for accurately detection of weak physical
signals and for predicting the values of a small number of physical variables
of interest based on a relatively large number of noisy feature signals. In
particular, we consider the situation where the feature signals can be
continuously measured at all times, but the weak signal and the target
variables can be assessed only in a special and well controlled environment
with certain additional measurements - the calibration phase. The question is,
in the deployment phase where the additional measurements for extracting the
weak signal are no longer available and the target variables are not accessible
any more, whether the weak signal and the variables can be predicted based on
the available feature signals?

A directly relevant application is magnetic navigation in a GPS (global
positioning system) denied environment, where the goal is to use the Earth's
anomaly magnetic field for precise positioning of an airplane, as this field
is position-dependent~\cite{canciani2016absolute,canciani2021magnetic}.
In such an application, various sensors in the cockpit of the the airplane are
employed to generate a large number of feature signals for extracting the
Earth's anomaly magnetic field through complex mathematical algorithms. With
a predetermined map between the anomaly field and the position for the flying
region, if the anomaly field can be accurately detected, precise
positioning can be achieved. A key challenge is that the anomaly magnetic
field is weak and the feature signals from the sensors are overwhelmingly
noisy due to the extensive electronic equipment in the cockpit and the other
(dominant) components of the Earth's magnetic field. The problem then becomes
one of detecting a weak signal from a noisy background that can be several
orders of magnitude stronger than the signal and determining the components of
the instantaneous position vector of the airplane (the target variables).
In this regard, recently a machine-learning scheme based on reservoir 
computing or time-delayed feed-forward neural networks was developed to
detect the Earth's anomaly field, with the implied equivalent positioning 
accuracy in the range of 10 to 40 meters~\cite{zhai2023detecting}.

Here, we articulate a random-forest based machine-learning scheme to detect the 
Earth anomaly magnetic field and simultaneously to determine the position of the 
flying airplane. Different from the previous work~\cite{zhai2023detecting}, 
we do not combine machine learning with other widely used calibration methods
such as the standard Tolles-Lawson (TL) model. Rather, in the whole process,
only the machine-learning model is employed to generate the anomaly field 
signal {\em and} the positioning information based on the input feature 
signals. From time-series data gathered from the cockpit of an airplane,
we demonstrate that the random-forest algorithm performs remarkably well in 
filtering Earth's anomaly magnetic field and generating the instantaneous 
position of the aircraft. With the aid of the conventional inertial navigation 
system (INS), the positioning error can be reduced to less than 10 meters with 
negligible standard deviation or uncertainty. We also find that, contrary to 
the conventional wisdom, the classic TL model for calibrating and removing the 
magnetic field generated by the body of the aircraft is not necessary and may 
even be detrimental for the success of the random-forest method.

We remark that, for the task of precise positioning, our proposed random-forest
method in fact does not require the use of INS sensors, for three reasons. First, 
the INS sensors have a limited range, making it impossible for them to be used 
for long-distance navigation. Second, due to the accumulation of minute 
inaccuracies in the readings, the INS sensors are subject to drift errors that 
reduce the accuracy over time. Third, connecting INS sensors with other 
navigation systems may be difficult and complex. 

We wish to emphasize the dynamical nature of the complex signals in our study.
The measured signal comprises various components, including the weak signal
generated by the Earth's anomaly magnetic field (the target signal to be
detected), the signal from the other (dominant) components of the Earth's
magnetic field, and the signals generated by the electronic equipment within
the airplane cockpit. While the signals other than the target signal represent
some kind of ``noises'' to be removed and are typically much stronger, they are
in fact complex signals with their own dynamics and time scales. To correctly
describe these signals that need to be removed, we use the term ''strong
complex signals'' or ``overwhelming complex signals.'' It is the dynamical
nature of these strong signals to be removed which makes the random-forest
approach effective. We note a recent work demonstrating that noises in the
conventional sense can be filtered out by machine-learning schemes such as
reservoir computing~\cite{nathe2023reservoir}.

In Sec.~\ref{sec:background}, we provide a brief overview of the background 
of magnetic navigation and the TL model for flight magnetic-field calibration 
and some previous machine-learning methods. In Sec.~\ref{sec:methods}, we 
describe the flight datasets, machine-learning methods employed in our study
(for the purpose of performance comparison), simulation and data-preprocessing 
details. Section~\ref{sec:results} presents results on feature selection, 
detection of the anomaly magnetic field, and precise positioning. A discussion 
and potential future research conclude the paper in Sec.~\ref{sec:discussion}.

\section{Background} \label{sec:background}

\subsection{Earth's anomaly magnetic field for navigation in a GPS-denied environment} \label{subsec:anomaly_field}

The Earth's magnetic field, or the geomagnetic field, is comprised of several 
field components~\cite{gnadt2020signal}. The main component is the core field 
generated by the motion of molten iron in the Earth's outer core, with its 
magnitude ranging from 25 to 65 micro-Tesla at the surface of the Earth. While
the core field makes compasses point north and is responsible for geophysical 
phenomena such as the auroras, its magnitude is still quite weak: about 25-65 
micro-Tesla at the surface of the Earth, which is about 100 times weaker than 
a refrigerator magnet. The second component is the crustal anomaly field 
generated by the Earth's crust and upper mantle, whose magnitude is about 100 
nano-Tesla, which is about 100 times weaker than the core field. While the core
field is the dominant component of the geomagnetic field, it is weakly 
time-dependent and is not sensitive to changes in the position, rendering it 
unsuitable for precise positioning and navigation. In contrast, the anomaly
field is position-dependent and has much stronger spatial variations than the 
core field. Consequently, in principle it is possible to use exploit the 
anomaly field for navigation. 

The widely used GPS can achieve the positioning 
accuracy of less than 10 m worldwide. However, because GPS signals are weak 
electromagnetic signals and must be transmitted over long distances, it is 
vulnerable to external interference such as jamming or 
spoofing~\cite{ioannides2016known}. In a GPS-denied environment, alternative 
navigation systems are needed for positioning, which include radio-based 
navigation~\cite{kayton1997avionics}, computer-vision-based 
navigation~\cite{veth2006fusion}, star-trackers~\cite{liebe2002accuracy}, 
terrain height matching~\cite{chen2015review}, and gravity 
gradiometry~\cite{richeson2007gravity}. Despite the outstanding performance 
of these navigation approaches in some specific scenarios, they are unable to
work universally under different circumstances. For instance, terrain-aided 
navigation relies on the unique features of the terrain, which will lose 
efficacy when working around oceans and deserts, and star-trackers rely on 
the stars, so it is not workable during the day or in cloudy weather. 
Different from these methods, the Earth's anomaly field is approximately 
time-invariant but strongly spatially variant, making magnetic navigation 
an appealing alternative~\cite{canciani2016absolute,canciani2021magnetic} to
GPS. Indeed, the anomaly-field based magnetic navigation is limited neither to 
terrains now to the time of the day. Another advantage is that, unlike active 
navigation such as GPS, magnetic navigation is a kind of passive navigation 
and, due to the power of the magnetic field decreasing as $d^{-3}$ 
with distance $d$, it is not practically possible to disrupt a magnetic 
navigation device through jamming~\cite{mandea2010geomagnetic}. A great 
challenge of magnetic navigation is that the anomaly field is extremely weak 
and usually is embedded in an overwhelmingly strong noisy background. For example, 
in the cockpit of a flying aircraft, various types of electronic devices are 
in active operation~\cite{erdmann2021geomagnetic}. To make magnetic navigation 
feasible, extracting the weak anomaly-field signal from strong complex signal 
is essential. With the availability of a predetermined magnetic map, the extracted 
clean anomaly-field signal can be used to determine the instantaneous position of 
the aircraft, possibly with the aid of a standard INS~\cite{gnadt2022machine}.

\subsection{Tolles-Lawson model} \label{subsec:TL_ML}

To realize magnetic navigation, effective methods to extract the anomaly
magnetic-field signal and to obtain the real positioning information are 
needed. For a flying airplane, the magnetic field generated by the body of 
the aircraft must be removed from the measured signal to yield the Earth's
magnetic field. The Tolles-Lawson (TL) 
model~\cite{tolles1950magnetic,tolles1954compensation,han2017modified} is a
linear aeromagnetic compensation method that estimates the magnetic field 
generated by the aircraft from the total measured magnetic field. When the 
aircraft is flying ideally in a ``magnetically quiet'' mode, e.g., there are 
only limited radio transmissions~\cite{gnadt2022advanced}, the TL model 
performs well in extracting the anomaly field from the signals measured by
magnetometers placed on the exterior surface of the airplane. However, when 
the magnetometers are placed inside the cockpit of the airplane, the TL model 
will not be sufficient to remove the overwhelming complex signals. 
In spite of this, the TL model still represents a state-of-the-art model to 
calibrate the anomaly field through magnetometers placed outside the airplane 
and for pre-filtering the data. 

\subsection{Previous machine-learning methods}

The last thirty years have witnessed the use of machine learning for magnetic 
navigation. Earlier, neural networks were proposed~\cite{williams1993aeromagnetic} 
as a model-free method for aeromagnetic calibration. About three years ago, 
hundreds complicated neural networks were trained and it was 
demonstrated~\cite{hezel2020improving} that the nonlinear machine-learning 
method is capable of reducing the external added noise and extracting the 
magnetic anomaly signal. It was proposed~\cite{gnadt2022machine} recently 
that the anomaly field can be extracted with small errors by combining the TL 
model and machine learning. In particular, an extended Kalman filter was used to 
demonstrate that the extracted anomaly field can lead to low positioning errors. 
More recently, two machine-learning methods, one based on recurrent neural 
networks and another using feed forward neural networks with time-delayed inputs, 
in combination with the TL model, were articulated~\cite{zhai2023detecting} for 
detecting the weak anomaly field from measurements performed inside the cockpit 
of a flying airplane.

\section{Methods} \label{sec:methods}

\begin{figure} [ht!]
\centering
\includegraphics[width=1\linewidth]{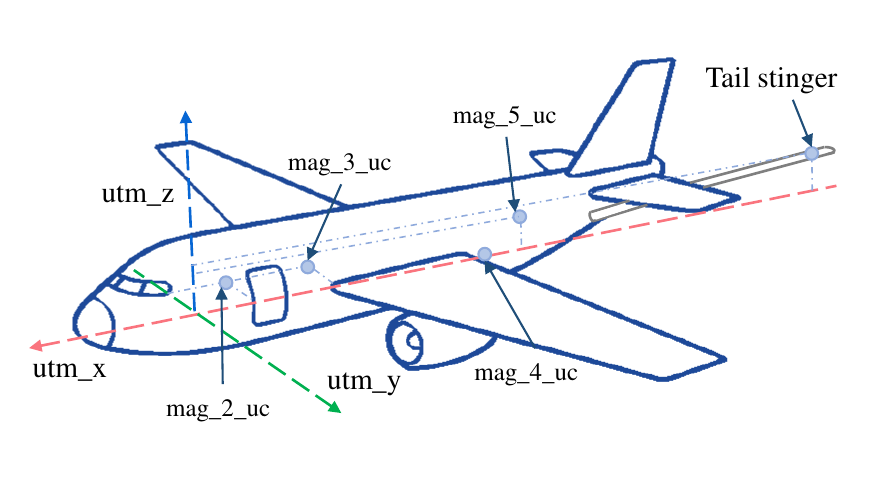} 
\caption{Configuration of magnetometers on the aircraft. The signals from the 
magnetometers inside the airplane contain strong complex signals produced by the 
various electronic devices. The signal from the magnetometer placed at the tail 
stinger is free from these overwhelming complex signals which, after the TL 
calibration, leads to the true magnetic anomaly field signal. The measurements 
from the other four magnetometers contain the anomaly field embedded in strong 
complex signals. The datasets are from test flights conducted by Sanders Geophysics 
Ltd. (SGL) near Ottawa, Canada.}
\label{fig:aircraft}
\end{figure}

\begin{figure*}
\centering
\subfloat[\centering Random forest - an ensemble of decision trees]
{{\includegraphics[width=8cm]{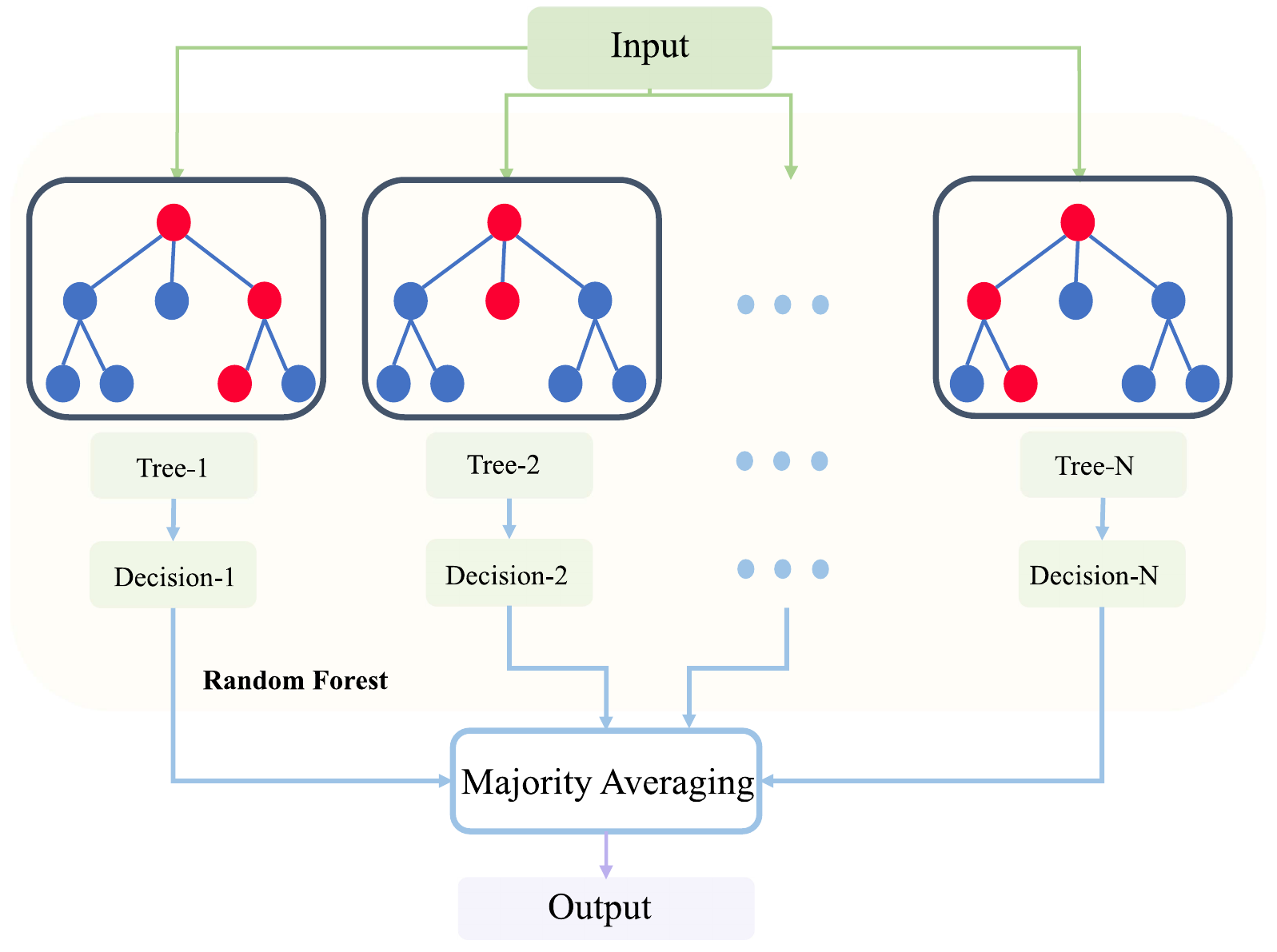} }}
\qquad
\subfloat[\centering An in depth look at a sample decision tree in a random 
forest]{{\includegraphics[width=8cm]{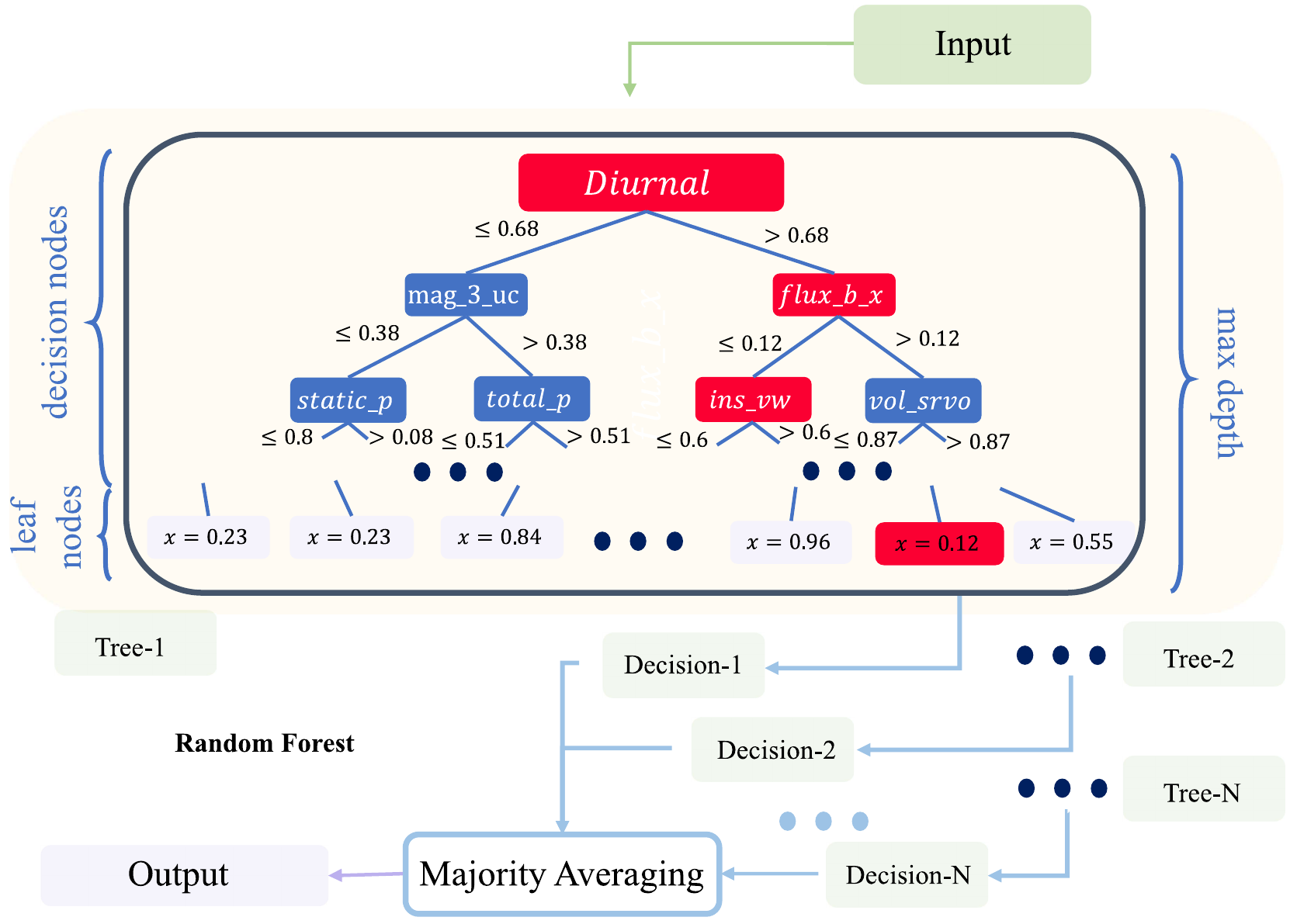} }}
\caption{A schematic illustration of a random forest of decision trees. 
(a) A random forest and (b) a decision tree in the ``forest.'' Prediction of the 
target variable of interest is achieved by taking the average of median of the 
predictions from all the trees in the forest.}
\label{fig:randomforest}
\end{figure*}

\subsection{Data source}

The datasets used in our work come from the signal enhancement for the magnetic 
navigation challenge problem~\cite{gnadt2020signal}. The goal was to extract a 
``clean'' magnetic anomaly field from measured complex signals by using a trained 
neural network. To achieve this, several test flights were conducted, each 
containing several segments (or lines). Five magnetometers were used to record 
the magnetic-field signals, one placed at the tail stinger of the aircraft (in 
a ``magnetically quiet'' mode), as shown in Fig.~\ref{fig:aircraft}. In 
particular, the tail-stinger signal was calibrated by the TL model, resulting in 
the ground truth of the true magnetic anomaly field. In our study, GPS signals 
were also used to provide the positioning information, representing the 
``ground truth'' of the aircraft position. 

\subsection{Three machine-learning methods used in this study}

The problem to extract the magnetic anomaly signal from noisy measurements belongs 
to signal filtering. Classical signal-processing methods such as linear filters or 
wavelet transformations are not suitable because the frequency bands of the embedded 
signal and strong complex signals overlap completely. Machine learning provides a 
potent and automated way of signal filtering, making it feasible to extract useful 
information from complex signals. In general, neural networks can be powerful for 
processing complicated signals with nonlinear properties. For example, convolutional 
neural networks (CNNs)~\cite{o2015introduction,li2021survey,kim2017convolutional} have 
been used in signal filtering tasks including denoising electroencephalography 
(EEG) signals or removing motion artifacts from magnetic resonance imaging (MRI) 
data~\cite{QEZELBASHCHAMAK2022100418}. Here, we describe three machine-learning
methods used in our work.

\paragraph*{K-nearest neighbor (KNN) approach:}
The KNN approach~\cite{peterson2009k} is a non-parametric method used for 
regression and classification. The value of an instance in KNN regression is 
determined by the mean or median of the $k$ closest instances. The user 
selecting the parameter $k$ establishes how many nearest neighbors should be 
utilized to produce a forecast. This enables the method to model nonlinear 
connections and capture complex patterns in data without making any assumptions
about the distribution of the underlying data - a key benefit for certain 
problems~\cite{goyal2014suitability,kohli2021sales,song2017efficient}. The KNN 
method has several limitations, such as sensitivity to the choice of the distance 
metric, curse of dimensionality, and high computational complexity. To address 
these limitations, the decision-tree and random-forest methods can be used.

\paragraph*{Decision tree:}
A decision tree is also a regressor and classifier that requires iteratively 
segmenting the feature space into regions and constructing a tree-like 
structure performing prediction for each smaller 
portion~\cite{su2006fast,freund1999alternating}. The feature space is 
recursively split into two halves and a tree is constructed using the features 
that provide the largest sum-squared error reduction. The resulting structure 
is a binary tree, with the leaves designating the final regions and the projected
value being the mean or median of the target variable for the data points in 
that region. One advantage of decision tree regression is interpretability 
since the resulting tree structure can be visualized for understanding.
Decision trees can also handle nonlinear and multi-dimensional data and are 
robust to noise and outliers. Yet, decision tree regression is susceptible to
overfitting when the tree becomes complex and the data are noisy. To overcome 
the overfitting problem, pruning, early pausing, and regularization can be used to 
improve the generalization performance of a decision tree. Decision tree 
regression has been applied in a variety of fields, including engineering, 
finance, and environmental sciences, for applications such as building energy 
consumption modeling and air pollution forecasting~\cite{ramya2019fuzzy,yu2010decision,luo2021extraction,wu2006effective,nair2010decision}. 
 
\paragraph*{Random forest:}
As shown in Fig.~\ref{fig:randomforest}, a random forest is an ensemble of 
decision trees~\cite{breiman2001random,rigatti2017random,cutler2012random}, 
each trained using a different subset of characteristic features and data points. 
Techniques such as weighted voting or stacking can also be employed for 
constructing a random forest. A key benefit of a random forest is its ability 
to prevent overfitting since the ensemble of trees balances out the noise and 
variability in the data. Random forest is also resilient against noise and 
outliers. However, when there are too many or too few trees in the ensemble, a 
random forest may suffer from bias and correlation and may not work well with 
data that has complex dependencies or nonlinear interactions.

\subsection{Simulation hardware and software}

Our simulations were carried out on a desktop system with one NVIDIA GeForce 
GTX 750 Ti GPU, an Intel Core i7-6850K CPU @ 3.60GHZ, and 128 GB of RAM. 
During the training process, the $n\_jobs$ keyword (from the $model.fit$ 
method) was set to $-1$ so the simulations were done using all $32$ available 
logical cores. All codes were written in Python, where we used $sklearn$ - a 
machine learning python package - to train and test our algorithms. 

\subsection{Data preprocessing} \label{subsec:preprocessing}

We use real data selected from several flights (number $1002$ to $1007$)  
conducted by Sanders Geophysics Ltd. (SGL) near Ottawa, Canada. For instance, 
the dataset of flight $1002$ consists of $207580$ instances with the sampling 
time $dt=0.1$s, each comprised of 102 features from a collection of various 
sensor measurements, which are voltage, current, magnetic and other sensors 
as well as the positions, INS, and avionics systems readings. The position of 
the aircraft is derived from WGS xyz coordinates included in the dataset which 
are GPS positions and is the predicted target of the model, while the other 
features [from selected features or those from a principal-component analysis
(PCA)] are used as the inputs to the random-forest models. For random-forest 
models, the number of estimators (trees) is selected to be $100$. The 
dataset is used to train the models in order to filter the position out of the 
available sensor data. The performance is evaluated using the predicted 
root-mean square errors (RMSEs) on a head-out unseen test set. 

The original data contained missing values, outliers, and other anomalies,
rendering necessary extensive pre-processing. In general, machine learning 
methods require normalizing the data as an essential pre-processing 
step~\cite{bolstad2003comparison}. A simple method is to normalize the dataset
so that all the features have zero mean and unit variance. Another method is 
minmax, where each feature is scaled individually such that it falls in a given 
range of the training set, e.g., between zero and one. Normalization in signal 
filtering also helps in removing bias in the signal that may be brought on by 
variations in scale or units among the features. In our case, normalizing the 
data will ensure that any two features with different scales, e.g., one measured 
in nano Tesla and the other in Ampere, have the same scale. We find that, for our 
datasets, the minmax scalar method outperforms the standard scalar normalization. 
Because of the remaining variance after minmax scaling, it is useful to remove 
features with low variance since they contribute little to the process of 
filtering~\cite{alam2011study}. This can be done by setting a proper threshold 
value of the variance. Removing the features with variances below the threshold 
brings additional benefits such as speeding up feature selection, reducing 
overfitting, and making the model and results more interpretable. 
Table~\ref{tab:var} sorts the features of the normalized flight data (including 
flight number $1002$ to $1007$) in terms of their standard deviation. As a 
reasonable assumption, the exclusion variance threshold is set as $0.0025$. For 
the flight data, this means that the features $cur\_flap$, $mag\_2\_uc$, and 
$cur\_com\_1$ are removed at this step.  

\begin{table}[b]
\caption{\label{tab:var}
Standard deviations of the features of entire flight dataset}
\begin{ruledtabular}
\begin{tabular}{cc|cc}
feat. with lowest std & std & feat. with highest std & std\\
\hline
$cur\_flap$ &  0.0077 & $total\_p$ & 0.3438 \\
$mag\_2\_uc$ & 0.0282 & $cur\_ac\_lo$ & 0.3345\\
$cur\_com\_1$ & 0.0441 & $static\_p$ & 0.3344 \\
$ins\_acc\_z$ & 0.0545 & $baro$ & 0.3340 \\
$nrml\_acc$ & 0.0569 & $ins\_alt$ & 0.3323 \\
$roll\_rate$ & 0.0590 & $utm\_z$ & 0.3322 \\
$ltrl\_acc$ & 0.0715 & $msl$ & 0.3321 \\
$pitch\_rate$ & 0.0718 & $diurnal$ & 0.3123 \\
$mag\_1\_igrf$ & 0.0738 & $ins\_vw$ & 0.3063 \\
$cur\_srvo\_o$ & 0.0798 & $ins\_vn$ & 0.2963 \\
\end{tabular}
\end{ruledtabular}
\end{table}

\section{Results} \label{sec:results}

\subsection{Feature selection}

\begin{figure*}[ht!]
\centering
\includegraphics[width=0.75\linewidth]{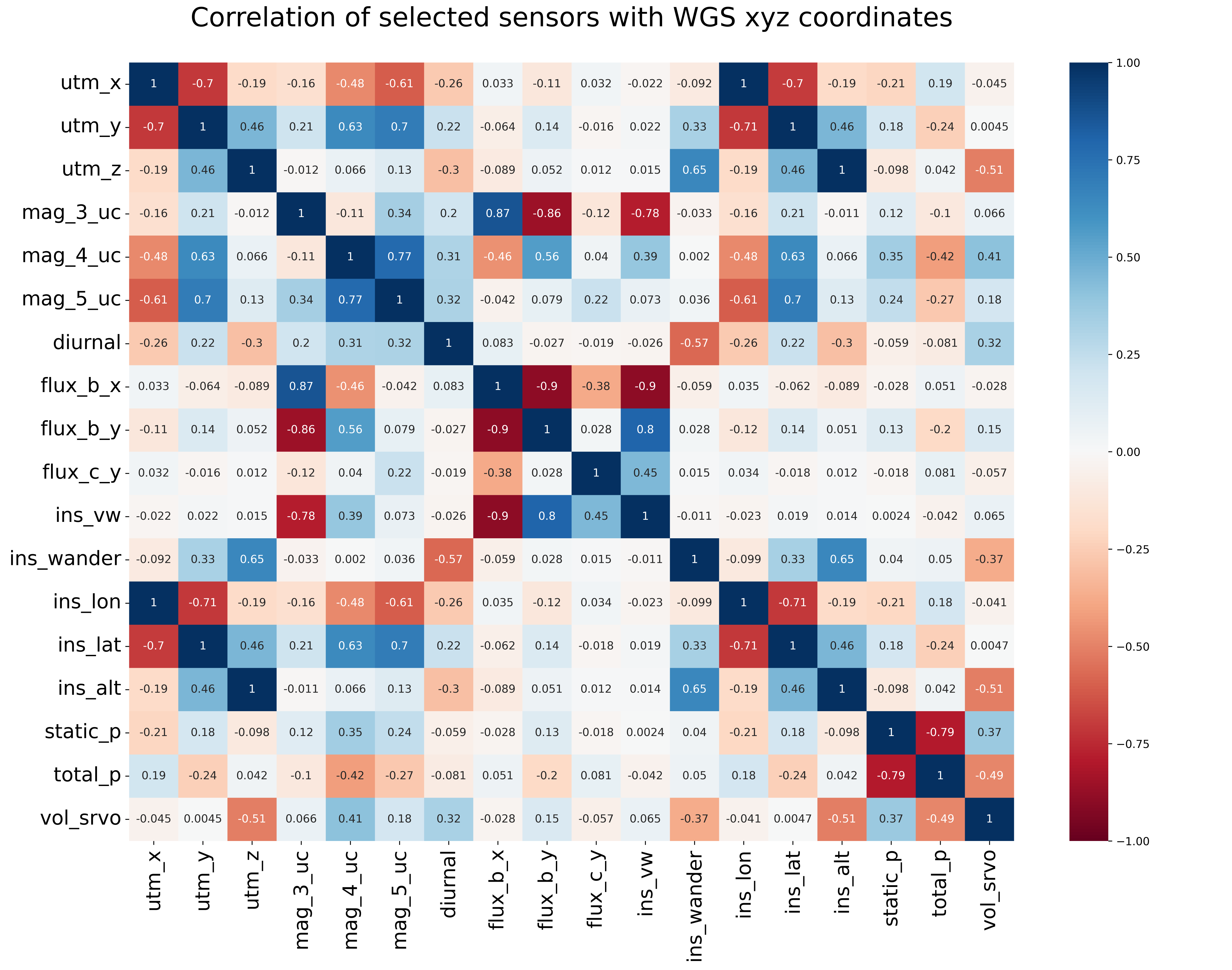} 
\caption{Correlation among the selected features in Table~\ref{tab:selfeat}. 
The information from the correlation is used to prune the redundant features. High
correlation among the features can produce ``multicollinearity,'' where numerous 
independent variables in a model are interrelated, making it challenging to 
predict the individual impact of each feature on the dependent variable.} 
\label{fig:corr}
\end{figure*}

A key step is sequential feature selection in which features are added to or 
removed from the features set in accordance with a predefined criterion such as 
RMSE, $R2$, or $F1$ score (depending on the application) until an optimal subset
of features is found~\cite{aha1995comparative}. Here, optimality means finding 
the smallest set of features leading to the required predictive performance,
which not only makes the model more understandable, but also reduces 
overfitting, increases prediction accuracy, and results in a computationally 
efficient training process. Given a set of features, optimal selection can be 
performed forward, backward, or a mix of both. Because of the relatively large
number of features in our dataset, we exploit the forward selection algorithm, 
which means that, at each step, the algorithm selects the best feature to add 
or remove based on the cross-validation score of the trained random-forest 
model. Our computation leads to 12 features: $mag\_3\_uc$, $mag\_4\_uc$, 
$mag\_5\_uc$, $diurnal$, $flux\_b\_x$, $flux\_b\_y$, $flux\_c\_y$, $ins\_vw$, 
$ins\_wander$, $static\_p$, $total\_p$, and $vol\_srvo$. Note that the 
$ins\_lon$, $ins\_lat$, and $ins\_alt$ features are excluded from the pool 
because they are equivalent to the target positions, as shown in 
Fig.~\ref{fig:corr}. For the INS free method, the $ins\_vw$ and $ins\_wander$
features are removed. Moreover, for the TL aided model, the 
$mag\_3\_uc$, $mag\_4\_uc$, and $mag\_5\_uc$ features are replaced by 
$mag\_3\_c$, $mag\_4\_c$, and $mag\_5\_c$, which are the compensated values 
of the magnetometers after application of the TL method. The details 
and description of the employed features for each method are listed in 
Tab.\ref{tab:selfeat}. 

\begin{table} [ht!]
\caption{\label{tab:selfeat}
Selected features using forward sequential feature selection on the entire flight dataset}
\begin{ruledtabular}
\begin{tabular}{c|c||c|c|c}
feat. & Description & \begin{tabular}{@{}c@{}}TL INS \\ free\end{tabular} & \begin{tabular}{@{}c@{}}INS \\ aided\end{tabular} & \begin{tabular}{@{}c@{}}TL INS \\ aided\end{tabular}\\
\hline
$mag\_3\_uc$ &  uncomp. mag. sensor $3$ &  \cmark & \cmark & \xmark \\
$mag\_4\_uc$ & uncomp. mag. sensor $4$ & \cmark & \cmark & \xmark \\
$mag\_5\_uc$ & uncomp. mag. sensor $5$ & \cmark & \cmark & \xmark \\
$mag\_3\_c$ & comp. mag. sensor $3$ & \xmark & \xmark & \cmark \\ 
$mag\_4\_c$ & comp. mag. sensor $4$ & \xmark & \xmark & \cmark \\
$mag\_5\_c$ & comp. mag. sensor $5$ & \xmark & \xmark & \cmark \\
$diurnal$ & measured diurnal & \cmark & \cmark & \cmark \\
$flux\_b\_x$ & fluxgate B  x-axis & \cmark & \cmark & \cmark \\
$flux\_b\_y$ & fluxgate B y-axis & \cmark & \cmark & \cmark \\
$flux\_c\_y$ & fluxgate C y-axis & \cmark & \cmark & \cmark \\
$ins\_vw$ & INS west velocity & \xmark & \cmark & \cmark \\
$ins\_wander$ & INS wander angle & \xmark & \cmark & \cmark \\  
$static\_p$ & avionics static pressure & \cmark & \cmark & \cmark \\
$total\_p$ & avionics total pressure & \cmark & \cmark & \cmark \\
$vol\_srvo$ & volt. sensors: servos & \xmark & \cmark & \cmark \\
\end{tabular}
\end{ruledtabular}
\end{table}

\begin{figure*}
\centering
\subfloat[\centering The first two components of the PCA transformation of 
the entire flight dataset in a 2D view. ]
{{\includegraphics[width=8cm]{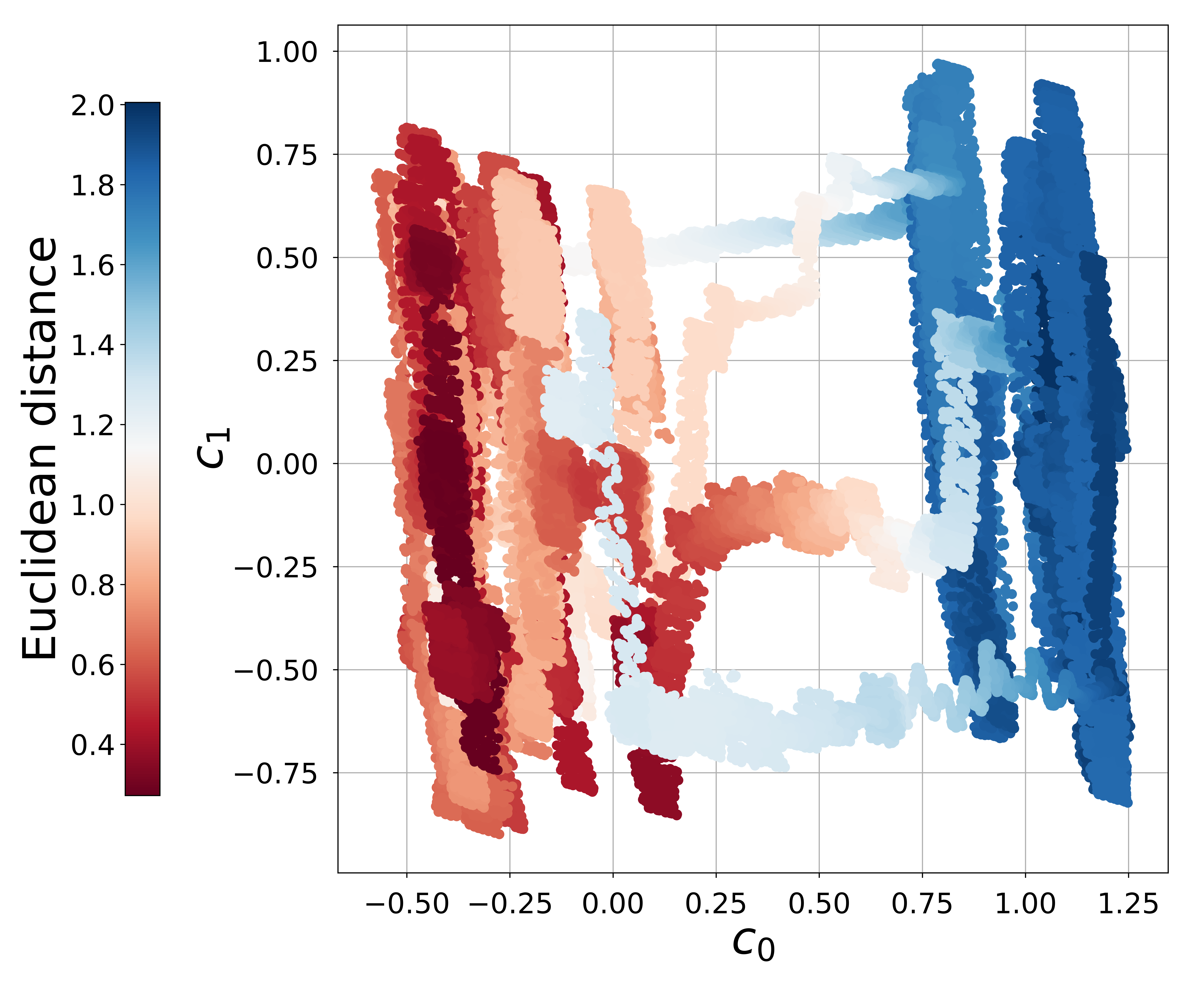} }}
\qquad
\subfloat[\centering The first three components of the PCA transformation of 
the entire flight dataset in a 3D view.]
{{\includegraphics[width=8cm]{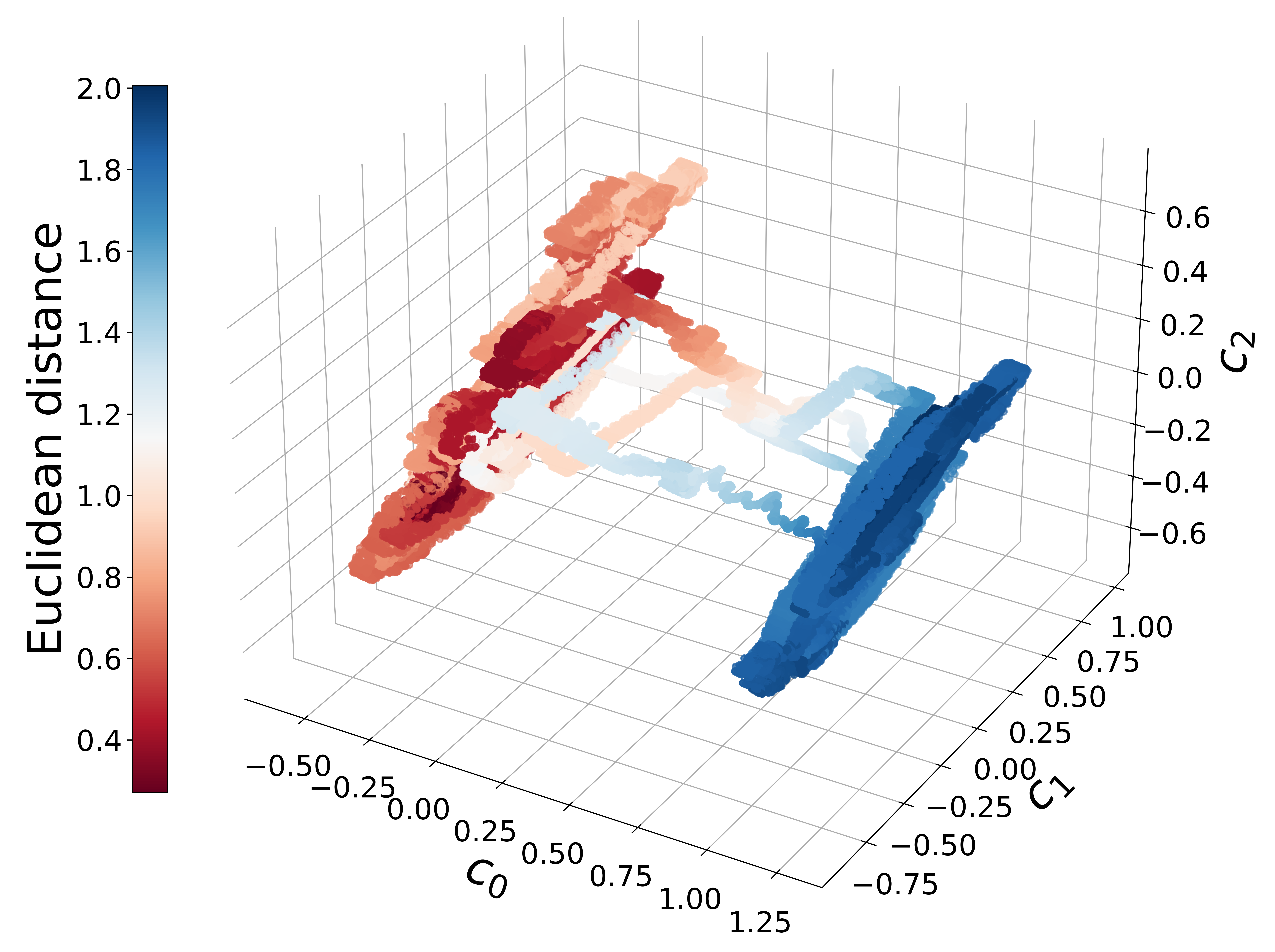} }}
\caption{Colored representation of the normalized Euclidean distance of the 
dataset instances from the origin. Distinct clusters in the data in relative 
Euclidean distance can be observed.}
\label{fig:pca}
\end{figure*}

We apply PCA to further process the feature data, where the components are
sorted by the eigenvectors in the order of their corresponding eigenvalues
from the highest to the lowest. There are two reasons for the PCA analysis. 
First, for the selected features to be effective, the correlations among them 
cannot be too high nor too low. In particular, high correlations can produce 
``multicollinearity,'' where numerous independent variables in a model are 
interrelated, making it difficult to predict the individual impact of each 
feature on the dependent variable. Figure~\ref{fig:corr} displays the 
correlation among the selected features, where it can be seen that the features
associated with the INS sensors have high correlations. For our magnetic
navigation problem, the degree and direction of the association among the 
sensor readings and the aircraft position are affected by feature correlations.
The highly correlated features can be integrated into new, uncorrelated features
named principal components through PCA for dimension reduction, which can then 
be used as inputs to our machine-learning based filter to remove the 
strong complex signals and 
redundant data. Second, PCA can improve visualization and help better understand
the data by transforming the original high-dimensional data into low-dimensional
ones. To establish a proper trade-off between dimensionality reduction and 
information retention, we consider the number of main components beyond which 
feature reduction can lead to information loss. Figures~\ref{fig:pca}(a) 
and \ref{fig:pca}(b) show the first two and three components of PCA applied to 
the set of selected features (without TL and INS), respectively, where the 
color elements represent the normalized Euclidean distance of the dataset 
instances from the origin. It can be seen from the PCA components that there 
exist distinct clusters in the data in terms of the Euclidean distance. 

We remark that nonlinear feature reduction methods such as Isomap and 
Kernel PCA are not suitable for our problem because of the large sizes of 
the datasets. In particular, to compute the kernel matrix of size 
$(data\_samples,data\_samples)$, it is necessary to compute and store 
$data\_samples^2$ number of terms. For our datasets, this requires about 
$312$GB of computer RAM. A potential solution is to perform clustering on 
the dataset and fill the kernel with the means of those clusters. However, 
even this approach might produce a large kernel matrix.

\subsection{Random-forest based detection of weak anomaly magnetic field}

\begin{figure}[]
\centering
\includegraphics[width=1\linewidth]{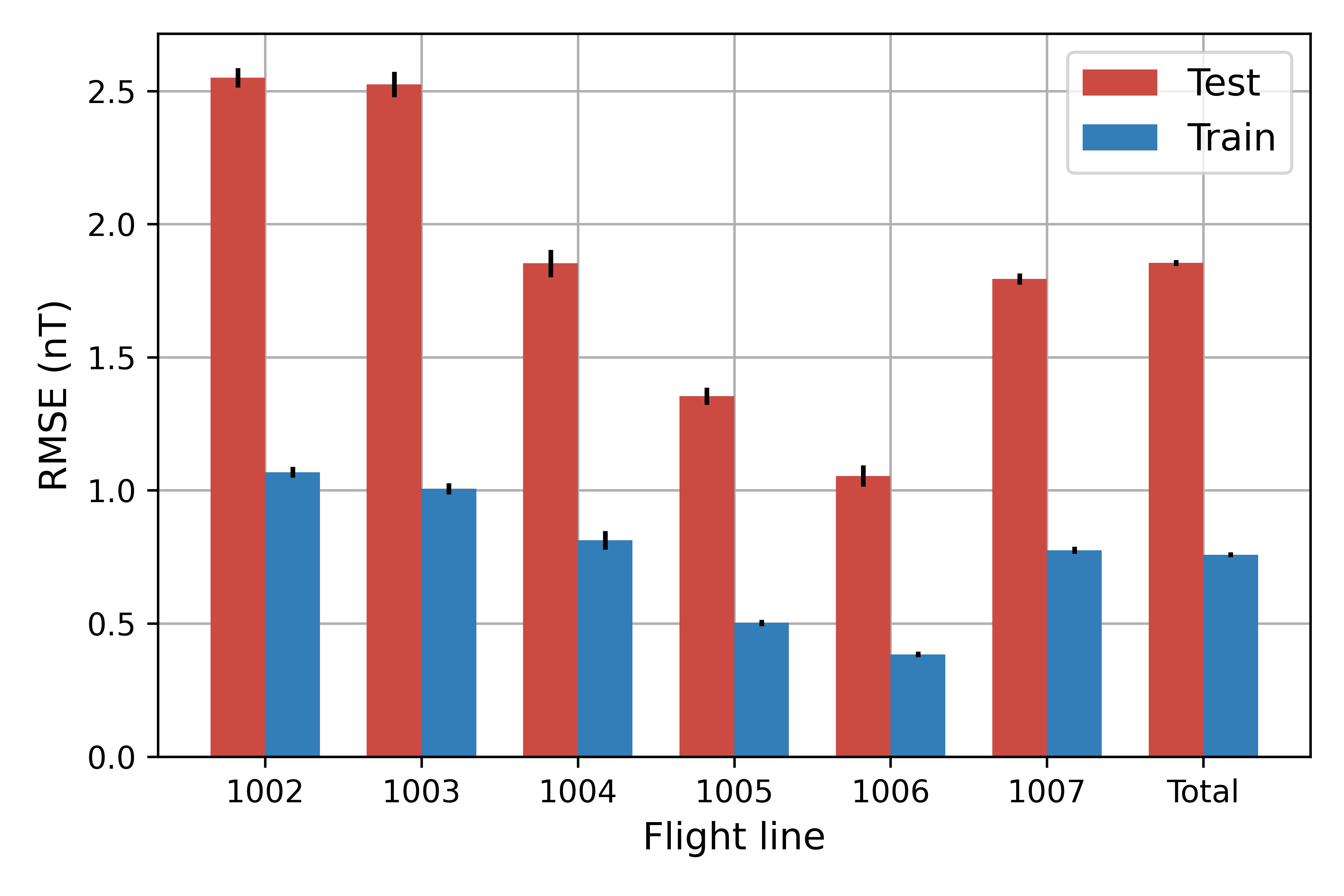} 
\caption{Results of random-forest based detection of weak signals. Selected 
features are used to detect the weak anomaly magnetic field signal. Compared
with a recent work on this topic based on the machine-learning methods of 
reservoir computing and feed forward neural networks ~\cite{zhai2023detecting},
the average RMSEs from the random-forest method are reduced by over 100\%.}
\label{fig:slg}
\end{figure}

We demonstrate the power of our random-forest method to detect weak anomaly 
field from data. For comparison, we also display the results from the KNN and 
decision-tree methods. Each machine-learning model uses the chosen optimal set 
of 12 features as input and the magnetic anomaly field signal as the output. To 
ensure a fair comparison, we perform hyperparameter tuning and calculate the 
average RMSE of the magnetic anomaly field for different flight lines. 
Figure~\ref{fig:slg} shows that, for our random-forest model, the average test 
RMSE is about $1.9$ nT for all flights. Table~\ref{tab:comp_slg} lists the average 
RMSEs from the different machine-learning methods. It can be seen that the KNN 
method has the largest error, the decision tree method suffers from overfitting 
but it is overcome by the random-forest method. These results indicate that 
random forests is a reliable approach for detecting weak magnetic anomaly fields 
from data. 

\begin{table*}[]
\caption{\label{tab:comp_slg}
Performance comparison among KNN, decision-tree, and random-forest methods in 
terms of RMSEs using selected features to detect weak anomaly magnetic field 
signal (in units of nT)}
\begin{ruledtabular}
\begin{tabular}{c||cccccc}
flight line & 
\begin{tabular}{@{}c@{}}Random Forest \\ train\end{tabular} & 
\begin{tabular}{@{}c@{}}Random Forest \\ test\end{tabular} & 
\begin{tabular}{@{}c@{}}KNN \\ train\end{tabular} & 
\begin{tabular}{@{}c@{}}KNN \\ test\end{tabular} & 
\begin{tabular}{@{}c@{}}Decision Tree \\ train\end{tabular} & 
\begin{tabular}{@{}c@{}}Decision Tree \\ test\end{tabular}
\\
\hline
1002 & 1.05 & 2.57 & 40.94 & 53.23 & 1.92 & 5.01\\
1003 & 1.05 & 2.60 & 19.29 & 25.68 & 1.17 & 3.94\\
1004 & 0.82 & 1.83 & 15.43 & 19.98 & 0.36 & 3.33\\
1005 & 0.50 & 1.31 & 9.46 & 12.74 & 0.58 & 3.16\\
1006 & 0.36 & 0.95 & 12.29 & 16.35 & 0.91 & 2.23\\
1007 & 0.80 & 1.84 & 18.18 & 23.34 & 1.27 & 3.68\\
all & 0.93 & 2.58 & 26.24 & 33.79 & 4.01 & 6.10\\
\end{tabular}
\end{ruledtabular}
\end{table*}

\subsection{Random-forest based precise positioning}

For navigation positioning, to ensure that the appropriate features are selected,
we carry out a feature-importance analysis. For a tree based method such as 
decision tree and random forests, it is practically impossible to analyze the 
feature importance by the model weights. Feature importance analysis is crucial 
for decision trees and random forests. Traditional metrics such as Gini 
impurity~\cite{disha2022performance} and mean decrease in 
impurity~\cite{han2016variable} are often used. 
Combining these techniques with domain expertise and cross-validation can lead
to a more comprehensive understanding of feature importance. We use two standard 
methods: permutation and dropping~\cite{kaneko2022cross}. In features permutation,
the performance of the model is assessed by randomly permuting the values of a 
single feature and comparing the results with a baseline model: the more the 
performance is degraded, the more important that feature is. In the 
feature-dropping approach, the model's performance is assessed by eliminating one 
feature at a time and monitoring how the result changes. Each feature's importance
is deduced from the performance drop that results from the removal of that certain
feature. These methods are computationally efficient for large datasets, and they 
serve to evaluate the significance of each feature without any knowledge about 
the intrinsic dynamics of the trained machine learning model, making them suitable
for a variety of applications. 

\begin{figure}[]
\centering
\includegraphics[width=1\linewidth]{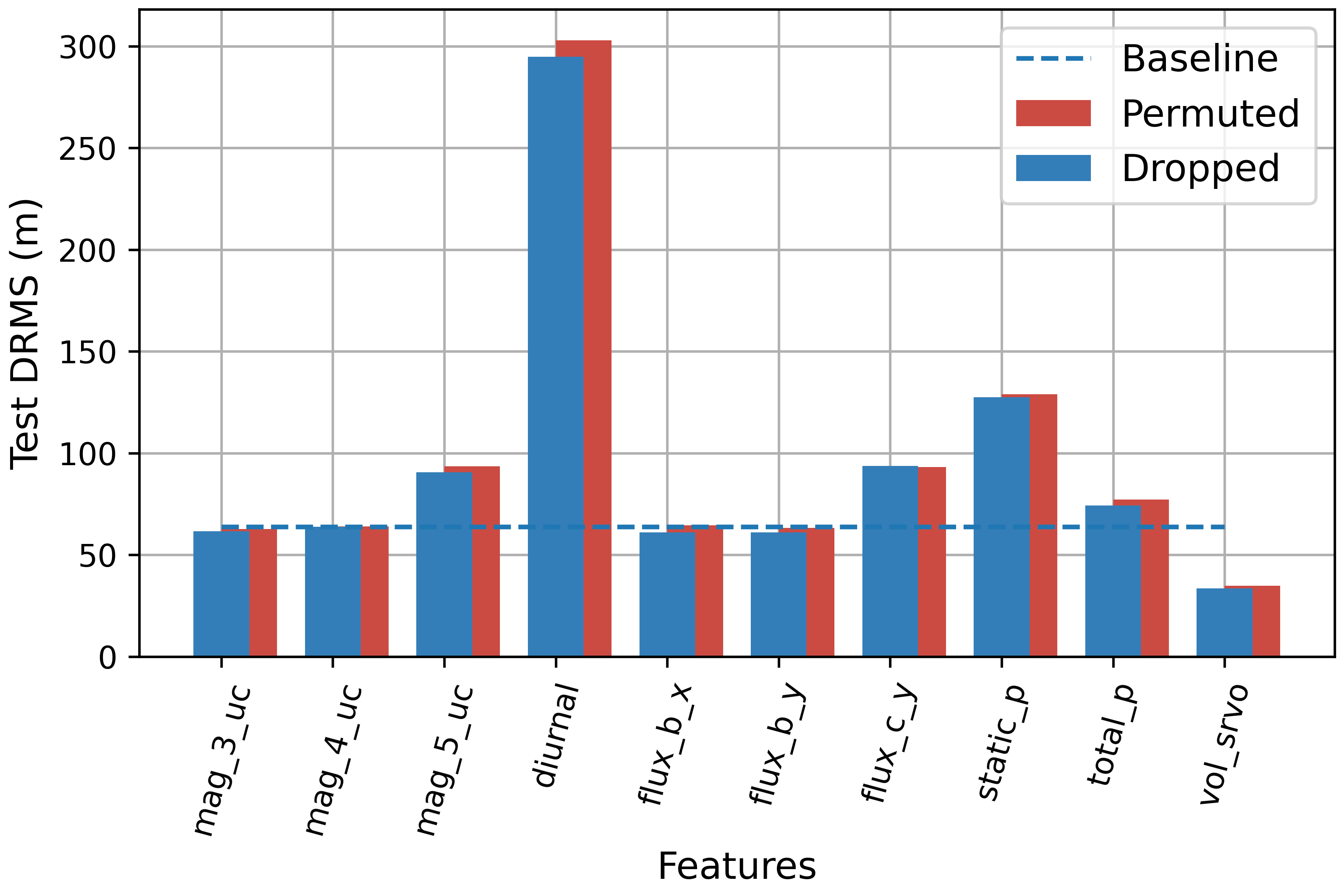} 
\caption{Importance analysis of selected features in the absence of INS. Random 
forest filtering of the position of the aircraft using the selected features in 
Tab.~\ref{tab:selfeat} is performed on a testing dataset. Both the permutation 
and dropping methods give that the feature $diurnal$ is the most significant. 
Other selected features are less significant, and removing the $vol\_srvo$ 
feature can improve the performance.}
\label{fig:imp1}
\end{figure}

\begin{figure}[]
\centering
\includegraphics[width=1\linewidth]{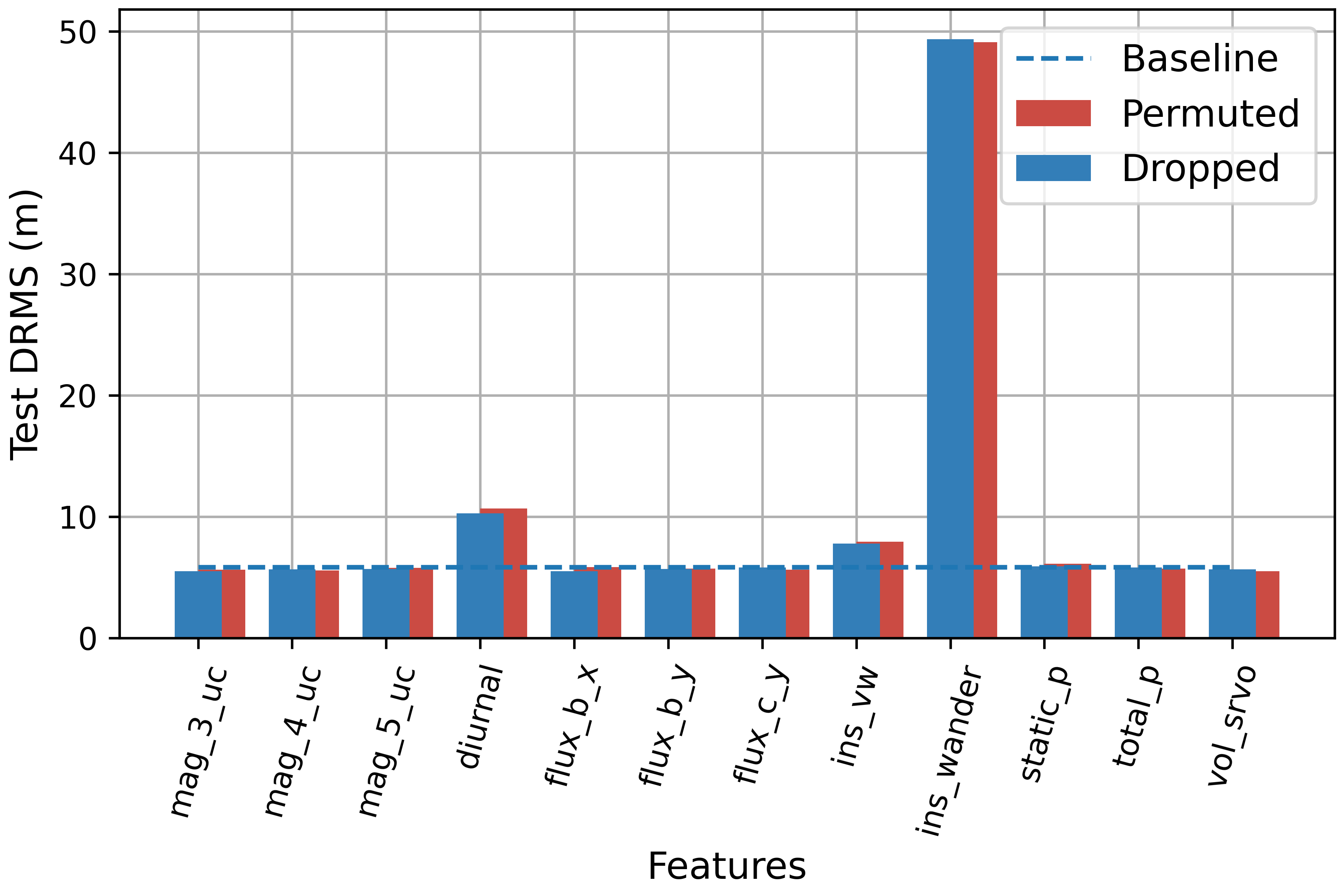} 
\caption{Importance analysis of selected features with INS. The position of the 
aircraft was filtered using a random-forest algorithm on a test set utilizing 
the selected features in Tab.~\ref{tab:selfeat}. Both the permutation and 
dropping methods give that the $ins\_wander$ feature is the most significant.
The second most significant feature is $diurnal$.}
\label{fig:imp2}
\end{figure}

Figures~\ref{fig:imp1} and \ref{fig:imp2} illustrate the feature importance 
analysis while performing the random-forest filtering of the position of the 
aircraft using the selected features in Tab.~\ref{tab:selfeat}. It can be 
seen that, for the INS free case, both the permutation and dropping methods
give that the $diurnal$ feature is the most significant. Note that the $diurnal$ feature can only be measured independently with an external base-station and won't generally be available on the aircraft. There are some approaches to modeling it based on past history, or using a statistical approach that could be used. For the INS aided model, the $ins\_wander$ feature is the most important, since the performance 
degrades significantly in its absence compared to the baseline method. 

\begin{figure}[ht!]
\centering
\includegraphics[width=\linewidth]{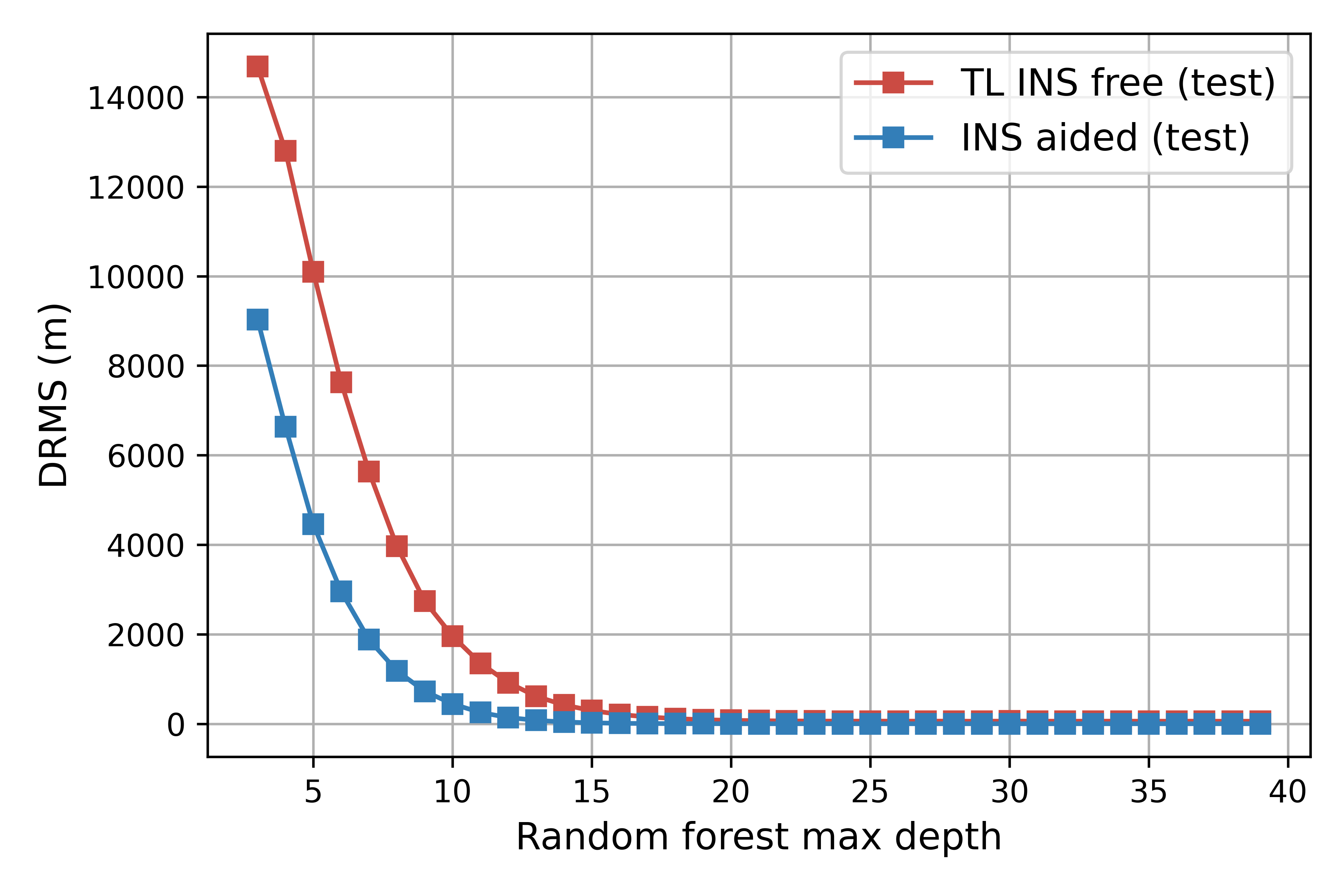} 
\caption{Determining the hyperparameter for random forests. Shown is the RMSE 
versus $max\_depth$, the depth of the forest (single hyperparameter). The 
optimal value of $max\_depth$ is around $25$.}
\label{fig:depth}
\end{figure}

We compare the performance of three methods: KNN, decision tree, and random 
forests, in positioning using both the selected features (Tab.~\ref{tab:selfeat})
and PCA features in terms of RMSEs from training and testing sets on different 
flight datasets. An important initial step is to determine the hyperparameter
values. In general machine learning, the hyperparameters are those that cannot 
be learned from the data but must be pre-determined before training, the
selection of which is critical to the success of machine learning. For the 
decision tree and KNN methods, we use PyGad~\cite{suganuma2017genetic} to tune 
the hyperparameter values. The random-forest method has a single hyperparameter
- $max\_depth$. Figure~\ref{fig:depth} shows the RMSE versus $max\_depth$, which 
gives that the optimal value of $max\_depth$ is around $25$. 

\begin{figure}[]
\centering
\includegraphics[width=1\linewidth]{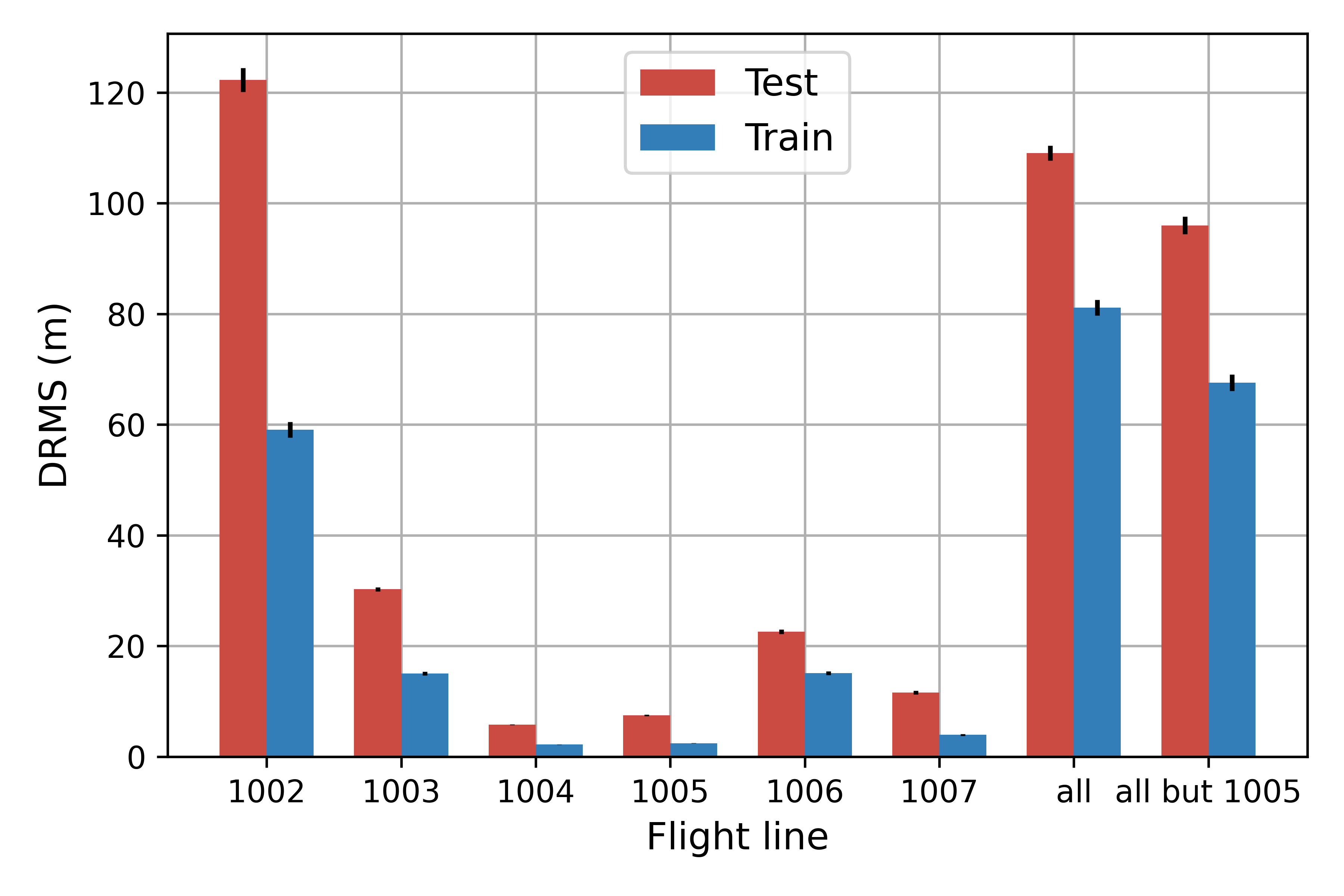} 
\caption{Results from the TL INS method. The compensated values from the 
magnetic sensors, the Tolles-Lawson model, and INS sensors are used. The 
comparison between the TL INS free and INS aided methods suggests that including
the TL model tends to degrade the positioning performance.} 
\label{fig:tlins}
\end{figure}

\begin{figure}[]
\centering
\includegraphics[width=1\linewidth]{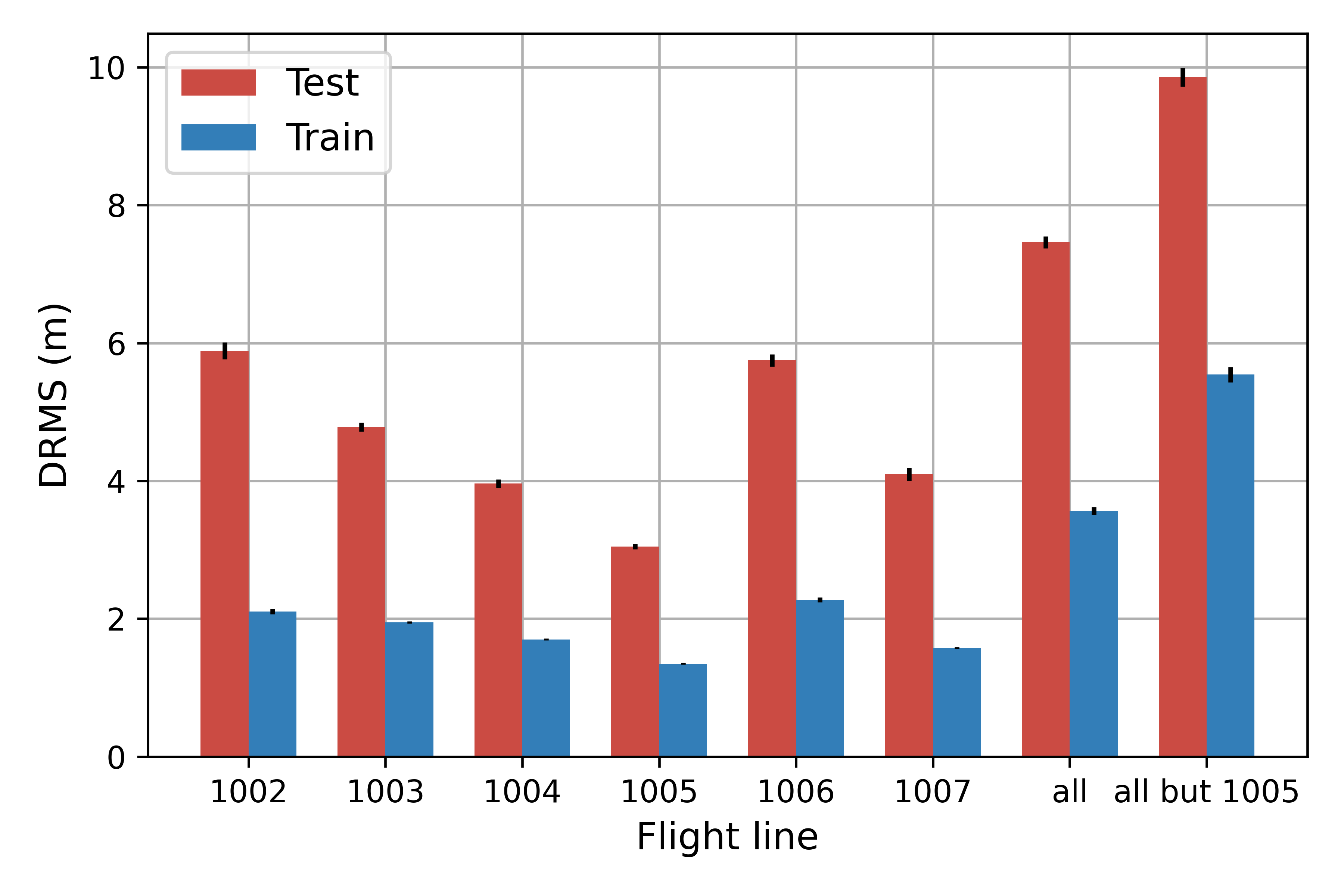} 
\caption{Results from the INS aided method. The uncompensated values from the
magnetic sensors and INS sensors are used. The INS aided method gives the best 
performance.}
\label{fig:ins}
\end{figure}

\begin{figure}[]
\centering
\includegraphics[width=1\linewidth]{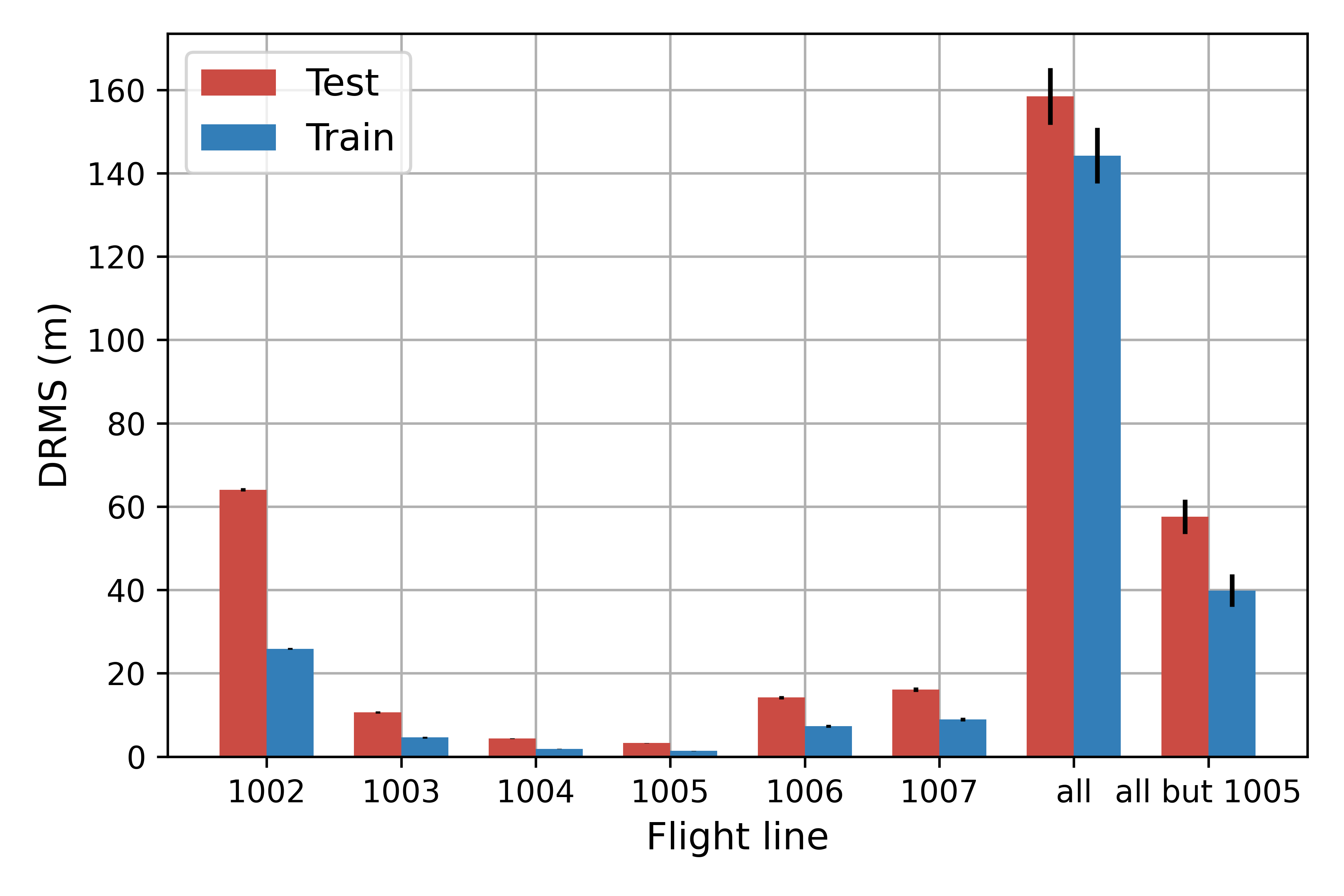} 
\caption{Results from the INS free method. The uncompensated values from 
magnetic sensors using the Tolles-Lawson model and INS sensors are not used. 
While the INS aided method gives the best performance, the INS free method 
does not rely on TL model or INS sensors.}
\label{fig:insfree}
\end{figure}

Table~\ref{tab:comp} summarizes the best-performance results with INS data in 
terms of DRMS, the distance between the actual and predicted positions of the 
aircraft. The $all$ flight line means that the entire dataset in the training 
and testing is used, whereas $all\ but\ 1005$ denotes that the entire dataset 
except that from flight number $1005$ is used for training but the test is 
performed on the entire dataset including flight number $1005$. The results show 
that in all cases, the selected features perform better than the PCA features. 
Remarkably, random forests outperforms the other two methods in the testing DRMS. 
For the decision-tree method, overfitting is pronounced because it performs better 
than random forests in training but worse in testing. Figure~\ref{fig:tlins} 
shows the results of the compensated values from magnetic sensors from the 
TL model and INS sensors. Figures~\ref{fig:ins} and \ref{fig:insfree} display the 
DRMS results when INS data are used and excluded, respectively. There are 
significant variations in the errors across the three methods. For example, 
the INS aided method consistently performs better than the other two methods, 
while the TL INS free aided method outperforms the TL INS aided method. An
implication is that incorporating INS sensors generally leads to improved 
accuracy. For all three methods, the test errors are consistently lower than 
the training errors, indicating that the models are capable of generalization
from training to test data. The INS aided method can be a reliable alternative 
that provides good accuracy with less reliance on a specific model. The INS free
methods, while less accurate, are suitable for applications where INS sensors 
are unavailable or too costly to implement. Another advantage of the INS free 
method is that it is independent of TL model.

Table~\ref{tab:total} summarizes the positioning results in terms of the mean 
and standard deviations from the entire experiment for different methods. A
comparison between the TL INS free and INS aided methods indicates that using 
the TL model tends to degrade the positioning performance, while the INS aided 
method gives the best performance. 

\begin{table*} [ht!]
\caption{\label{tab:comp}
Performance comparison in terms of DRMS among KNN, decision tree, and 
random-forest methods using PCA and selected features for different flight 
datasets using the INS aided method (in units of meters)}
\begin{ruledtabular}
\begin{tabular}{ccc||cccccc}
flight line & 
use PCA & 
\begin{tabular}{@{}c@{}}use selected \\ features\end{tabular} & 
\begin{tabular}{@{}c@{}}Random Forest \\ train\end{tabular} & 
\begin{tabular}{@{}c@{}}Random Forest \\ test\end{tabular} & 
\begin{tabular}{@{}c@{}}KNN \\ train\end{tabular} & 
\begin{tabular}{@{}c@{}}KNN \\ test\end{tabular} & 
\begin{tabular}{@{}c@{}}Decision Tree \\ train\end{tabular} & 
\begin{tabular}{@{}c@{}}Decision Tree \\ test\end{tabular}
\\
\hline
1002 & \xmark & \cmark & 2.12 & 6.00 & 30.62 & 42.14 & 1.34 & 8.50\\
1003 & \xmark & \cmark & 1.94 & 4.75 & 37.21 & 53.11 & 2.46 & 9.61\\
1004 & \xmark & \cmark & 1.70 & 3.82 & 33.15 & 47.63 & 0.68 & 7.76\\
1005 & \xmark & \cmark & 1.35 & 3.07 & 19.83 & 29.11 & 0.13 & 7.19\\
1006 & \xmark & \cmark & 2.02 & 5.48 & 25.85 & 36.78 & 13.52 & 20.84\\
1007 & \xmark & \cmark & 1.55 & 4.05 & 38.89 & 53.26 & 0.41 & 7.71\\
all & \xmark & \cmark & 2.54 & 6.51 & 115.96 & 159.9 & 48.48 & 53.90\\
all but 1005 & \xmark & \cmark & 2.74 & 7.02 & 103.17 & 125.93 & 73.25 & 82.06\\
1002 & \cmark & \xmark & 726.59 & 2370.4 & 869.7 & 1277.05 & 745.92 & 2924.68\\
1003 & \cmark & \xmark & 252.08 & 2355.1 & 417.7 & 2506.68 & 466.07 & 2760.00\\
1004 & \cmark & \xmark & 171.21 & 738.1 & 264.7 & 577.04 & 121.29 & 946.71\\
1005 & \cmark & \xmark & 213.53 & 840.0 & 515.2 & 742.15 & 42.10 & 1107.38\\
1006 & \cmark & \xmark & 75.48 & 347.5 & 139.4 & 217.58 & 10.24 & 415.45\\
1007 & \cmark & \xmark & 246.63 & 1759.5 & 524.7 & 1823.29 & 233.97 & 1830.46\\
all & \cmark & \xmark & 663.09 & 5324.7 & 1075.9 & 5582.99 & 1835.13 & 6108.65\\
all but 1005 & \cmark & \xmark & 547.7 & 17501.4 & 783.73 & 547.75 & 1949.34 & 19254.25\\
\end{tabular}
\end{ruledtabular}
\end{table*}

\begin{table}[]
\caption{\label{tab:total}
Mean and standard deviation of RMSE from the entire experiment for different 
methods (in units of meters)}
\begin{ruledtabular}
\begin{tabular}{c|cccc}
Method & \begin{tabular}{@{}c@{}} Test \\ mean\end{tabular} & \begin{tabular}{@{}c@{}}Test \\ std\end{tabular} & \begin{tabular}{@{}c@{}}Train \\ mean\end{tabular} & \begin{tabular}{@{}c@{}}Train \\ std\end{tabular}\\
\hline
\hline
\begin{tabular}{@{}c@{}}TL INS \\ free\end{tabular} & 41.08 & 2.50 & 29.26 & 2.46 \\
\hline
\begin{tabular}{@{}c@{}}INS \\ aided\end{tabular} & 5.60 & 0.03 & 2.51 & 0.03 \\
\hline
\begin{tabular}{@{}c@{}}TL INS \\ aided\end{tabular} & 50.62 & 0.78 & 30.79 & 0.66 \\
\end{tabular}
\end{ruledtabular}
\end{table}

\section{Discussion} \label{sec:discussion}

Exploiting machine learning to detect weak physical signals immersed in 
strong complex signals has attracted a recent interest with 
applications such as extracting the
Earth's anomaly magnetic field from overwhelmingly noisy signals collected from
the magnetic sensors installed inside the cockpit of a flying airplane. Previously,
the specific machine-learning architectures explored for this application were 
reservoir computing (recurrent neural networks) and time-delayed feed forward 
neural networks~\cite{zhai2023detecting}. It was demonstrated that, when combined 
with the classical TL model for removing the aircraft magnetic field, the weak 
anomaly magnetic field can be reliably detected. Detecting the anomaly field has 
a direct application in magnetic navigation in a GPS-denied 
environment~\cite{canciani2016absolute,canciani2021magnetic}. For the navigation
problem, the goal is precise positioning: obtaining the instantaneous position of
the flying airplane, a problem that was not addressed in the recent 
work~\cite{zhai2023detecting}, raising the need for a more general approach
to magnetic navigation.

The present work develops a random-forest based machine-learning approach to
magnetic navigation, where the problems of detecting the weak anomaly magnetic 
field and determining the position of the flying airplane are solved 
simultaneously. When many feature signals are present, as in the flying data
available to us, a vast and complex ``random forest'' can be constructed and
trained. The random forest contains an enormously large number of distinct paths 
(from the ``root'' to the ``leafs''), each being associated with a specific value 
of the anomaly magnetic field and the position of the airplane. In the testing 
phase when the information about the anomaly field and the position is not 
available, from a set of features signals, a well-trained random forest can be 
efficiently searched to yield the anomaly field and the position. As many 
available feature signals contain redundant information, we performed a process 
to select the optimal set of feature signals so as to greatly improve the 
computational efficiency. As random forests is essentially an ``intelligent'' 
book-keeping machine, all required for training is the selected set of feature 
signals and the corresponding anomaly field and position, thereby removing the 
need of calibration methods, e.g., the TL model. Indeed, we demonstrated that 
high accuracies in the anomaly field and position can be achieved even without TL 
calibration.

More generally, the random-forest framework developed here can be applied to 
signal filtering, an important task in many fields including image processing, 
speech recognition, and economic forecasting. There are many challenges in 
designing effective filtering algorithms, such as dealing with noise, 
nonstationary signals, and nonlinear dependencies. The success of deploying 
random forests to magnetic navigation reported here can serve as a starting
point to generalize the machine-learning model to other applications. For the
broad task of signal filtering, a potential future research direction could be 
to investigate deep learning methods, such as convolutional neural networks, 
specially temporal graph convolutional neural networks~\cite{8809901}.
Additionally, transfer learning could be explored, where a pre-trained model 
is fine-tuned on a new dataset for signal filtering. In this case, the knowledge 
learned from previous flights can be employed in learning new flights data, 
thereby requiring shorter flight times. Furthermore, the development of online 
learning algorithms for signal filtering could be explored, which could adapt 
to changes in the signal over time.

\section*{Data and Code Availability}

The data and code are available at: https://github.com/AminMoradiXL/magnav

\section*{Acknowledgment}
This work was supported by AFOSR under Grant No.~FA9550-21-1-0438.


%
\end{document}